\documentclass[12pt]{article}

\usepackage[dvips]{graphicx}
\usepackage{epsf, subfigure}
\usepackage{psfig}
\usepackage[latin1]{inputenc}
\usepackage[T1]{fontenc}
\usepackage{amsmath}
\numberwithin{equation}{section}

\begin{document}

\title{\Large Integrable and superintegrable quantum systems in a magnetic 
field}

\author{\normalsize Josée Bérubé\\
\footnotesize Département de mathématiques et de statistique et Centre de 
recherche 
mathématiques\\
\footnotesize Université de Montréal, C.P. 6128, succ. Centre-Ville\\
\footnotesize Montréal, Québec\\
\footnotesize H3C 3J7\\
\footnotesize Canada\\
\footnotesize berube@crm.umontreal.ca
\and
\normalsize Pavel Winternitz\\
\footnotesize Département de mathématiques et de statistique et Centre de 
recherche mathématiques\\
\footnotesize Université de Montréal, C.P. 6128, succ. Centre-Ville\\
\footnotesize Montréal, Québec\\
\footnotesize H3C 3J7\\
\footnotesize Canada\\
\footnotesize wintern@crm.umontreal.ca}

\maketitle

\begin{abstract}
Integrable quantum mechanical systems with magnetic fields are constructed in 
two-dimensional Euclidean space. The integral of motion is assumed to be a first 
or second order Hermitian operator. Contrary to the case of purely scalar 
potentials, quadratic integrability does not imply the separation of variables 
in the Schrödinger equation. Moreover, quantum and classical integrable systems 
do not necessarily coincide: the Hamiltonian can depend on the Planck constant 
$\hbar$ in a nontrivial manner.
\end{abstract}

\pagebreak

\section[]{Introduction}

The purpose of this article is to study the integrability properties of a 
quantum particle moving in an external magnetic field. More specifically, we 
will consider the Schrödinger equation in a two-dimensional Euclidean space with 
the Hamiltonian

\begin{equation}
H=-\frac{\hbar^2}{2}(\partial_{x}^2+\partial_{y}^2)-\frac{i\hbar}{2}[A(x,y)
\partial_{x}+\partial_{x}A(x,y)+B(x,y)\partial_{y}+\partial_{y}B(x,y)]+V(x,y).
\label{eqH}
\end{equation}

The vector and scalar potentials $(A,B)$ and $V$ are to be determined from the 
requirement that the system should be integrable, i.e. a well-defined quantum 
mechanical operator $X$ should exist, that commutes with the Hamiltonian, i.e.

\begin{equation}
[H,X]=0.
\end{equation}

In this particular study, we shall restrict to the case when $X$ is a first or 
second order polynomial in the momenta. We shall be particularly interested in 
the case of superintegrable systems, when two independent operators, $X_{1}$ and 
$X_{2}$, commuting with the Hamiltonian exist. In general, $X_{1}$ and $X_{2}$ 
do not commute with each other, but together generate an algebra of operators, 
commuting with $H$.

In classical mechanics, integrable systems are of interest, because they have 
regular trajectories. Indeed, their motion is restricted to a torus in phase 
space. Superintegrable systems are even more regular. Trajectories are 
completely determinded by the values of the $2n-1$ integrals of motion. In 
particular, all bounded trajectories are periodic, as in the case of the 
harmonic oscillator, or Kepler problem.

In quantum mechanics, integrability, i.e. the existence  of $n$ integrals of 
motion, provides a complete set of quantum numbers, characterizing the system. 
Moreover, it simplifies the calculation of energy levels and wave functions. 
Superintegrability, in all cases studied so for, entails exact solvability. This 
means that energy levels in superintegrable systems can be calculated 
algebraicly, i.e. they satisfy algebraic rather than transcendental equations.

Previous searches for integrable and superintegrable systems in quantum 
mechanics concentrated on scalar potentials only~\cite{fris1}-\cite{rod}. It was 
established that for scalar potentials the existence of first and second order 
integrals of motion implies the separation variables in the Schrödinger 
equation, and also in the Hamilton-Jacobi equation in classical mechanics. 
Moreover, for scalar potentials and second order integrals of motion, classical 
and quantum integrable systems coincide (i.e. classical and quantum potentials 
are the same).

Surprisingly, when third order integrals are considered, a new phenomenon 
occurs: integrable and superintegrable quantum systems that have no classical 
counterpart~\cite{hiet1}-\cite{gravel2}. Indeed, in the classical limit 
$\hbar\to 0$ the potential vanishes, $V(x,y)\to 0$ and we obtain free motion.

Previous studies of integrability in magnetic fields were conducted in the 
framework of classical mechanics~\cite{gram},~\cite{mcsween}. It was established 
that the existence of second order integrals of motion in the presence of 
magnetic fields no longer implies the separation of variables. However, the 
integrals of motion were still classified into equivalence classes under the 
action of the Euclidean group and the highest order terms have the same form as 
in the case of a purely scalar potential.

In this paper we restrict ourselves to the two-dimensional Euclidean space 
$E(2)$, the Hamiltonian~\eqref{eqH} and to first, or second order integrals. In 
Section 2 we formulate the problem of finding the integrals of motion, first in 
the classical, then in the quantum case. We show that the determining equations 
in 
the two cases are the same for first order integrals of motion, not however for 
second order ones. Section 3 is devoted to first order integrals of motion. They 
are shown to exist if and only if the magnetic field and an effective scalar 
potential are invariant under either translations, or rotations. We also show 
that superintegrability with two (or more) first order integrals occurs only for 
a constant magnetic field and effective potential. In Section 4 we consider a 
specific class of second order operators which we call ``cartesian integrals''. 
In the absence of a magnetic field they lead to separation of variables in 
cartesian coordinates. We also show that superintegrability with one cartesian 
integral and a second integral of any (quadratic) type occurs only for a 
constant magnetic field. In the cartesian case there is no difference between 
classical and quantum integrability. Polar integrability and superintegrability 
are investigated in Section 5. All cases of integrability with one ``polar'' 
integral of motion are identified. The quantum case differs from the classical 
one and the magnetic field can depend on the Planck constant $\hbar$ in a 
nontrivial manner. In Section 6 we show that a polar integral can exist 
simultaneously with any other independent second order integral only if the 
magnetic field is constant. The final Section 7 is devoted to conclusions and 
open problems.

\section[]{Formulation of the problem}

\subsection{Classical mechanics}

Since we will be comparing results in quantum and classical mechanis, let us 
briefly recapitulate some results obtained earlier~\cite{gram},~\cite{mcsween}. 
The classical counterpart of the Hamiltonian~\eqref{eqH} is

\begin{equation}
H=\frac{1}{2}(p_{x}^2+p_{y}^2)+A(x,y)p_{x}+B(x,y)p_{y}+V(x,y),
\end{equation}

\noindent
where $p_{x}$ and $p_{y}$ are the momenta conjugate to $x$ and $y$, 
respectively. The classical equations of motion in the Hamiltonian form are

\begin{equation}
\begin{array}{ccc}
\dot{x}=\frac{\partial H}{\partial p_{x}}=p_{x}+A,~\dot{y}=\frac{\partial 
H}{\partial p_{y}}=p_{y}+B\\
\dot{p}_{x}=-\frac{\partial H}{\partial 
x}=-V_{x}-A_{x}p_{x}-B_{x}p_{y},~\dot{p}_{y}=
-\frac{\partial H}{\partial y}=-V_{y}-A_{y}p_{x}-B_{y}p_{y}\\
\end{array}
\label{eqmouv1}
\end{equation}

The equation of motion~\eqref{eqmouv1} can be rewritten in the Newton form as

\begin{equation}
\begin{array}{ccc}
\ddot{x}=-W_{x}+\Omega\dot{y},\\
\ddot{y}=-W_{y}-\Omega\dot{x},\\
\end{array}
\label{eqmouv2}
\end{equation}

\begin{equation}
\begin{array}{ccc}
W=V-\frac{1}{2}(A^2+B^2),\\
\Omega=A_{y}-B_{x}.\\
\end{array}
\end{equation}

The equations of motion~\eqref{eqmouv2} are invariant under a gauge 
transformation of the potentials

\begin{equation}
\begin{array}{ccc}
V(x,y)\to \tilde{V}(x,y)=V+(\vec{A},\bigtriangledown\phi)
+\frac{1}{2}(\bigtriangledown\phi)^2,\\
\vec{A}(x,y)\to \tilde{\vec{A}}(x,y)=\vec{A}+\bigtriangledown\phi,\\
\end{array}
\end{equation}

\noindent
where we have put $\vec{A}=(A,B)$ and $\phi=\phi(x,y)$ is an arbitrary smooth 
function. Thus, the quantities that are of actual physical importance are the 
magnetic field $\Omega$ and the effective potential $W$.

A classical first integral of motion is postulated to have the form

\begin{equation}
C=f_{1}(x,y)\dot{x}+f_{2}(x,y)\dot{y}+m(x,y).
\label{eqC}
\end{equation}

The determining equations for the funtions $f_{1}$, $f_{2}$ and $m$ are obtained 
from the requirement

\begin{equation}
\{ H,C\}
=\frac{dC}{dt}=0,
\label{poisson}
\end{equation}

\noindent
when eq.~\eqref{eqmouv1} are satisfied, i.e. $C$ is a constant on the solutions 
of the equations of motion and Poisson commutes with the Hamiltonian.

Similarly, a classical second order integral of motion has the form

\begin{equation}
C=g_{1}(x,y)\dot{x}^{2}+g_{2}(x,y)\dot{y}^{2}+g_{3}(x,y)\dot{x}\dot{y}+k
_{1}(x,y)\dot{x}+k_{2}(x,y)\dot{y}+m(x,y).
\label{eqC2}
\end{equation}

The determining equations for the functions $g_{i}$, $k_{i}$ and $m$ are again 
obtained from the condition~\eqref{poisson}.

The equations for the coefficients of the first and second order classical 
integrals of motion were derived and partially solved 
elsewhere~\cite{gram},~\cite{mcsween}. We shall give them again below as 
classical limits of the corresponding equations in the quantum case. To 
facilitate a comparison, we must rewrite the classical integrals in terms of 
momenta, rather than velocities, i.e. substitute $\dot{x}=p_{x}+A$, 
$\dot{y}=p_{y}+B$.

\subsection{Quantum mechanics}

In quantum mechanics an integral of motion will be a Hermitian operator $X$ that 
commutes with the Hamiltonian $H$.

Let us first consider a first order integral in the momenta:

\begin{equation}
X=-\frac{i\hbar}{2}(f_{1}\partial_{x}+\partial_{x}f_{1}+f_{2}\partial_{y}
+\partial_{y}f_{2})+f_{1}A+f_{2}B+m.
\label{eqX}
\end{equation}

The classical limit of the operator~\eqref{eqX} is the integral~\eqref{eqC}; 
$f_{1}$, $f_{2}$ and $m$ are functions of $x$ and $y$.

The commutator $[X,H]$ with $H$ as in eq.~\eqref{eqH} will contain second, 
first and zero order terms in the derivatives. Setting the coefficients of all 
of them equal to zero, we obtain the following set of determining equations 

\begin{equation}
f_{1,x}=0,~f_{2,y}=0,~f_{1,y}+f_{2,x}=0,
\label{eq123}
\end{equation}

\begin{equation}
-f_{2}\Omega+m_{x}=0,~f_{1}\Omega+m_{y}=0,~f_{1}W_{x}+f_{2}W_{y}=0.
\label{eq456}
\end{equation}

We see that the Planck constant $\hbar$ does not figure in eq.~\eqref{eq123} 
and~\eqref{eq456}. Hence these equations must coincide with their classical 
limit, and indeed, they do~\cite{gram}. In particular, eq.~\eqref{eq123} implies

\begin{equation}
f_{0}=\alpha y+\beta,~f_{1}=-\alpha x+\gamma,
\label{eqf1f2}
\end{equation}

\noindent
where $\alpha$, $\beta$ and $\gamma$ are real constants. Hence the leading terms 
(independent of $A$, $B$ and $m$) of the operator $X$ of eq.~\eqref{eqX} lie in 
the Lie algebra $e(2)$ of the Euclidean group $E(2)$, generated by

\begin{equation}
P_{1}=-i\hbar\partial_{x},~P_{2}=-i\hbar\partial_{y},~L_{3}=
-i\hbar(y\partial_{x}-x\partial_{y}).
\end{equation}

Thus, we have

\begin{equation}
X=\alpha L_{3}+\beta P_{1}+\gamma P_{2}+\alpha(yA-xB)+\beta A+\gamma B+m.
\label{Xo1}
\end{equation}

We shall write the second order operator corresponding to the 
integral~\eqref{eqC2}, after symmetrization, as

\begin{equation}
\begin{array}{ccc}
X=-\frac{1}{2}\hbar^{2}\{ 2g_{1}\partial_{x}^{2}+2g_{2}\partial_{y}^{2}
+2g_{3}\partial_{x}\partial_{y}\\
+(2g_{1,x}+g_{3,y})\partial_{x}+(2g_{2,y}+g_{3,x})\partial_{y}+g_{1,xx}+g_{2,yy}
+g_{3,xy}\}\\ 
-\frac{i\hbar}{2}\{(4g_{1}A+2g_{3}B+2k_{1})\partial_{x}+(4g_{2}B+2g_{3}A+2k_{2})
\partial_{y}\\
+2g_{1}A_{x}+2g_{1,x}A+2g_{2}B_{y}+2g_{2,y}B+g_{3}A_{y}+g_{3}B_{x}+Ag_{3,y}
+Bg_{3,x}\\+k_{1,x}+k_{2,y}\} 
+g_{1}A^2+g_{2}B^2+g_{3}AB+k_{1}A+k_{2}B+m\\
\end{array}
\label{opX}
\end{equation}

The commutativity condition $[H,X]=0$ implies the following set of determining 
relations:

\begin{equation}
g_{1,x}=0,~g_{2,y}=0,~g_{1,y}+g_{3,x}=0,~g_{2,x}+g_{3,y}=0,
\label{eqgi}
\end{equation}

\begin{equation}
\begin{array}{ccc}
k_{1x}-g_{3}\Omega=0,~k_{2y}+g_{3}\Omega=0,\\
2\Omega(g_{1}-g_{2})+k_{1y}+k_{2x}=0,\\
2g_{1}W_{x}+g_{3}W_{y}+k_{2}\Omega-m_{x}=0, \\
2g_{2}W_{y}+g_{3}W_{x}-k_{1}\Omega-m_{y}=0, \\
\end{array}
\label{eq2345}
\end{equation}

\begin{equation}
k_{1}W_{x}+k_{2}W_{y}+\frac{\hbar^{2}}{4}(g_{2x}\Omega_{y}-g_{1y}\Omega_{x})=0.
\label{eq6}
\end{equation}

Eq.~\eqref{eqgi} and~\eqref{eq2345} are the same as the classical 
ones~\cite{gram}. Eq.~\eqref{eq6} is however different. It involves the Planck 
constant and reduces to the classical case only in the limit $\hbar\to 0$. Thus, 
in the presence of a nonconstant magnetic field $\Omega(x,y)$, classical and 
quantum integrability differ!

Eq.~\eqref{eqgi} can be solved as in the classical case~\cite{gram} and they 
imply

\begin{equation}
\begin{array}{ccc}
g_{1}=\alpha y^{2}-\beta y+\delta, \\
g_{2}=\alpha x^{2}+\gamma x+\zeta,\\
g_{3}=-2\alpha xy+\beta x-\gamma y+\xi, \\
\end{array}
\label{gis}
\end{equation}

\noindent
where the Greek letters represent real constants. Substituting~\eqref{gis} 
into~\eqref{opX} we obtain the operator $X$ in the form

\begin{equation}
\begin{array}{ccc}
X=\alpha[(L_{3}+yA-xB)^2+\hbar^2]-\frac{1}{2}\beta[(L_{3}+yA-xB)(P_{1}+A)\\
+(P_{1}+a)(L_{3}+yA-xB)]-\frac{1}{2}\gamma[(L_{3}+yA-xB)(P_{2}+B)+(P_{2}+B)\\
(L_{3}+yA-xB)]+\delta(P_{1}+A)^2+\zeta(P_{2}+B)^2+\xi(P_{1}+A)(P_{2}+B)\\
-\frac{i\hbar}{2}(2k_{1}\partial_{x}+k_{1,x}+2k_{2}\partial_{y}+k_{2,y})
+k_{1}A+k_{2}B+m.\\
\end{array}
\label{opeX}
\end{equation}

Thus, the leading part of eq.~\eqref{opeX} lies in the envelopping algebra of 
$e(2)$. For $A=B=0$ this coincides with the case of a scalar 
potential~\cite{fris1},~\cite{fris2},~\cite{temp}. As in the scalar case we can 
simplify eq.~\eqref{opeX} by Euclidean transformations and linear combinations 
with the Hamiltonian. The operator $X$ is transformed into a similar operator, 
with new values of the constants $\alpha,...,\xi$. Four classes of such 
operators exist, represented by

\begin{equation}
\begin{array}{ccc}
X_{C}=(P_{1}+A)^2-\frac{i\hbar}{2}(2k_{1}\partial_{x}+k_{1,x}\\
+2k_{2}\partial_{y}+k_{2,y})+k_{1}A+k_{2}B+m,\\
\end{array}
\label{Xcart}
\end{equation}

\begin{equation}
\begin{array}{ccc}
X_{R}=(L_{3}+yA-xB)^2+\hbar^2-\frac{i\hbar}{2}(2k_{1}\partial_{x}\\
+k_{1,x}+2k_{2}\partial_{y}+k_{2,y})+k_{1}A+k_{2}B+m,\\
\end{array}
\label{Xpol}
\end{equation}

\begin{equation}
\begin{array}{ccc}
X_{P}=-\frac{1}{2}\{(L_{3}+yA-xB)(P_{1}+A)+(P_{1}+A)(L_{3}+yA-xB)\}\\
-\frac{i\hbar}{2}(2k_{1}\partial_{x}+k_{1,x}+2k_{2}\partial_{y}+k_{2,y})+k_{1}A+
k_{2}B+m,\\
\end{array}
\label{Xpar}
\end{equation}

\begin{equation}
\begin{array}{ccc}
X_{E}=(L_{3}+yA-xB)^2+\hbar^2+\sigma[(P_{1}+A)^2-(P_{2}+B)^2]\\
-\frac{i\hbar}{2}(2k_{1}\partial_{x}+k_{1,x}+2k_{2}\partial_{y}+k_{2,y})+k_{1}A+
k_{2}B+m,~\sigma>0.\\
\end{array}
\label{Xell}
\end{equation}

In the case of a purely scalar potential the existence of a commuting operator 
of the type $X_{C}$, $X_{R}$, $X_{P}$ or $X_{E}$ implies that the Schrödinger 
equation will allow separation of variables in cartesian, polar, parabolic or 
elliptic coordinates, respectively. In the last case $\sigma$ is related to the 
interfocal distance for the elliptic coordinates.

Substituting eq.~\eqref{gis} into eq.~\eqref{eq6} we obtain

\begin{equation}
k_{1}W_{x}+k_{2}W_{y}+\frac{\hbar^2}{4}[(2\alpha x+\gamma)\Omega_{y}
-(2\alpha y-\beta)\Omega_{x}]=0.
\label{eq225}
\end{equation}

Thus, for the special case $\alpha=\beta=\gamma=0$ classical and quantum 
integrability will coincide.

\section[]{First order integrability and superintegrability}

A first order integral~\eqref{eqX} in quantum mechanics will exist if the 
overdetermined system~\eqref{eq123} and~\eqref{eq456} has a solution. The 
general 
solution of eq.~\eqref{eq123} is given by eq.~\eqref{eqf1f2}. Substituting  
into~\eqref{eq456} we obtain:

\begin{equation}
(\alpha x-\gamma)\Omega+m_{x}=0,~(\alpha y+\beta)\Omega+m_{y}=0,~(\alpha 
y+\beta)W_{x}+(-\alpha x+\gamma)W_{y}=0.
\end{equation}

We are only interested in cases with a magnetic field present, i.e. $\Omega\ne 
0$. With no loss of generality, we need only distinguish two cases:

1. $\alpha=1$, $\beta=\gamma=0$.

We obtain

\begin{equation}
W=W(\rho),~\Omega=\Omega(\rho),~\rho=\sqrt{x^2+y^2}.
\end{equation}

We see that in this case the magnetic field $\Omega$ and the effective scalar 
potential must be spherically symmetric. The potentials in the 
Hamiltonian~\eqref{eqH} can by a gauge transformation be taken into

\begin{equation}
A=\int{\frac{\Omega(\rho)\rho 
d\rho}{\sqrt{\rho^2-x^2}}},~B=0,~V=W(\rho)+\frac{1}{2}A^2.
\end{equation}

The integral of motion is

\begin{equation}
X=L_{3}+y\int{\frac{\Omega(\rho)\rho d\rho}{\sqrt{\rho^2-x^2}}}
-\int{\rho\Omega(\rho)d\rho}.
\end{equation}

2. $\alpha=\beta=0$, $\gamma=1$.

The magnetic field  and effective potential are translationally invariant

\begin{equation}
\Omega=\Omega(x),~W=W(x),
\end{equation}

and we can take

\begin{equation}
\begin{array}{ccc}
A=\Omega y,~B=0,~V=W+\frac{1}{2}y^2\Omega^2,\\
X=P_{2}+\int{\Omega(x)dx}.\\
\end{array}
\end{equation}

The system~\eqref{eqH} will be first order superintegrable if at least two first 
order integrals~\eqref{Xo1} exist. This is only possible if the magnetic field 
and effective potential are constant:

\begin{equation}
\Omega=\Omega_{0},~W=W_{0}.
\end{equation}

In this case actually three operators commuting with the Hamiltonian exist and 
we have

\begin{equation}
H=\frac{1}{2}(i\hbar\partial_{x}-\Omega y)^2-\frac{\hbar^2}{2}\partial_{y}^2,
\end{equation}

\begin{equation}
X_{1}=P_{1},~X_{2}=P_{2}+x\Omega_{0},~X_{3}=L_{3}-\frac{1}{2}(x^2-y^2)
\Omega_{0},
\label{X123}
\end{equation}

where we have chosen the gauge to be such that

\begin{equation}
A=\Omega_{0}y,~B=0,~V=\frac{1}{2}\Omega^2y^2.
\end{equation}

The classical equations of motion~\eqref{eqmouv1} are easily solved. The 
trajectories are circles (and are hence all closed). The Schrödinger equation 
allows the separation of variables in cartesian coordinates. The solution is

\begin{equation}
\Psi(x,y)=e^{i\frac{\lambda}{\hbar}x}f(y),
\end{equation}

where $f(y)$ satisfies the harmonic oscillator equation

\begin{equation}
f''-\{(\frac{\Omega y+\lambda}{\hbar})^2-\frac{2E}{\hbar^2}\} f=0.
\end{equation}

The integrals of motion~\eqref{X123}, together with the constant $\Omega$, 
satisfy the commutation relations of a central extension of the Euclidean Lie 
algebra:

\begin{equation}
[X_{1},X_{2}]=-i\hbar\Omega_{0},~[X_{3},X_{1}]=-i\hbar X_{2},
~[X_{3},X_{2}]=i\hbar X_{1}.
\end{equation}

Only three of the integrals $X_{1}$, $X_{2}$, $X_{3}$ and $H$ can be independent 
and indeed they satisfy

\begin{equation}
X_{1}^2+X_{2}^2+2\Omega_{0}X_{3}-2H=0.
\end{equation}

In polar coordinates the Schrödinger equation

\begin{equation}
\begin{array}{ccc}
\{\frac{1}{2}(i\hbar\cos(\phi)\partial_{r}-\frac{\sin(\phi)}{r}\partial_{\phi}
-\Omega r\sin(\phi))^2\\
-\frac{1}{2}(\sin(\phi)\partial_{r}+\frac{\cos(\phi)}{r}
\partial_{\phi})^2\}\Psi=E\Psi\\
\end{array}
\end{equation}

R-separates~\cite{mcsween},~\cite{miller}, rather than separates, and we have

\begin{equation}
\Psi(r,\phi)=e^{-\frac{i}{4}\Omega r^2\sin 2\phi}J_{m}(kr)e^{im\phi},
~k^2=2E+m\Omega,
\end{equation}

where $J_{m}(kr)$ is a Bessel function.

\section[]{Cartesian integrability and superintegrability}

\subsection{Integrability}

In order to find integrable systems with a second order operator commuting with 
the Hamiltonian, we must solve the system~\eqref{eqgi} to~\eqref{eq6}. To do 
this, we first transform $X$ to its canonical form, i.e. one of~\eqref{Xcart} 
to~\eqref{Xell}. We start with the simplest case, namely $X_{C}$ 
of~\eqref{Xcart}. We call this the ''cartesian'' case, because for a purely 
scalar potential it corresponds to separation of variables in cartesian 
coordinates. It corresponds to $\alpha=\beta=\gamma=\zeta=\xi=0$ and $\delta=1$ 
in eq.~\eqref{opeX}. Eq.~\eqref{eq225} implies that the 
determining equations~\eqref{eqgi},~\eqref{eq2345} and~\eqref{eq6} are the same 
in the classical and quantum cases (the $\hbar^{2}$ term in eq.~\eqref{eq6} 
vanishes). For purely scalar potentials $\Omega=k_{1}=k_{2}=0$ we reobtain the 
known result $W=W_{0}(y)+m(x)$~\cite{fris1}. From now on we assume $\Omega\ne 
0$. For completeness, we reproduce the result obtained earlier~\cite{gram} in 
the classical case, since it is valid in the quantum case as well:

\begin{equation}
\begin{array}{ccc}
\Omega=f_{xx}+g_{yy},\\
W=\frac{a}{3}(g-f)^3-\frac{b+d}{2}(g-f)^2+(c+k-e)(g-f),\\
k_{1}=-g_{y},~k_{2}=-f_{x},\\
m=-\frac{a}{3}(g^3+2f^3-3gf^2)+b(fg-f^2)+\frac{d}{2}(g^2-f^2)+c(g-2f)+eg-kf.\\
\end{array}
\label{sol1}
\end{equation}

Here $a$, $b$, $c$, $d$, $e$ and $k$ are constants and the functions $f=f(x)$ 
and 
$g=g(y)$ satisfy 

\begin{equation}
\begin{array}{ccc}
f_{xx}=af^{2}+bf+c,~g_{yy}=-ag^{2}+dg+e,\\
f_{x}\ne 0,~g_{y}\ne 0.\\
\end{array}
\label{eq42}
\end{equation}

Two exceptional cases occur when we have $f_{x}=0$ or $g_{y}=0$. These however 
imply $\Omega=\Omega(x)$, $W=W(x)$ or $\Omega=\Omega(y)$, $W=W(y)$ respectively. 
Then a first order invariant exists and the second order one is simply its 
square.
The general solution of eq.~\eqref{eq42} are elliptic functions.

\subsection{Cartesian superintegrability}

We shall now assume that $\Omega$ and $W$ are such that one cartesian integral 
$X_{1}$ exists, i.e. they satisfy eq.~\eqref{sol1}. We require 
that a second integral $X_{2}$ of the type~\eqref{opeX} should exist, in 
addition to the considered cartesian one. We can simplify the integral $X_{2}$ 
by translation and by linear combinations with $X_{1}$ and $H$. Rotations can 
not be used, since they would change the form of the operator $X_{1}$ and of the 
Hamiltonian. Two cases must be considered, $\alpha\ne 0$ and $\alpha=0$.

Case 1: $\alpha\ne 0$

We set $\alpha=1$, by a translation we transform $(\beta, \gamma)\to (0,0)$, by 
linear combinations we set $(\delta,\zeta)\to (0,0)$. We are left with an 
operator $X_{2}$ in the form~\eqref{opeX} with $\alpha=1$, 
$\beta=\gamma=\delta=\zeta=0$. The constant $\xi$ and functions $k_{1}$, $k_{2}$ 
and $m$ must be determined from the system~\eqref{eq2345},~\eqref{eq6}. Let us 
consider the case when $\Omega$ and $W$ are as in eq.~\eqref{sol1}. The 
first two equations imply

\begin{equation}
\begin{array}{ccc}
k_{1}=-2y(xf_{x}-f)-x^2yg_{yy}+\xi f_{x}+\xi xg_{yy}+C_{1}(y),\\
k_{2}=xy^2f_{xx}+2x(yg_{y}-g)-\xi yf_{xx}-\xi g_{y}+C_{2}(x).\\
\end{array}
\end{equation}

We substitute $k_{1}$, $k_{2}$, $\Omega$ and $W$ into the remaining four 
equations and investigate their compatibility. After somewhat lengthy 
calculations we obtain a simple result: the equations are compatible for 
$\Omega\ne 0$ if and only if $\Omega$ and $W$ are constant. We arrive at the 
case (3.7), already investigated in section 3.

Case 2: $\alpha=0$

In order to obtain an independent second order integral we must have 
$\beta^2+\gamma^2\ne 0$ and we can normalize $\beta^2+\gamma^2=1$ and put 
$\delta=\zeta=\xi=0$ (by linear combinations with $H$ and $X_{1}$). The set of 
equations~\eqref{eqgi} to~\eqref{eq6}  is then again compatible only for 
$\Omega$ and $W$ constant.

The conclusion of this section is that for $\Omega\ne 0$ cartesian 
superintegrability 
with two second order integrals exists only in a trivial sense. Thus 
$\Omega=\Omega_{0}$, $W=W_{0}$ and all second order integrals are reducible: 
they are polynomials in the three first order ones.

\section[]{Polar integrability}

We now request that one second order integral should exist and that it be of the 
form~\eqref{Xpol}. We shall call this operator $X_{R}$ a polar type integral. 
Let us transform the determining equations~\eqref{eq2345},~\eqref{eq6} to polar 
coordinates $x=r\cos(\phi)$, $y=r\sin(\phi)$. The resulting equations are

\begin{equation}
P_{r}=0,~P+Q_{\phi}=0,
\label{pol1}
\end{equation}
\begin{equation}
2r^{3}\Omega-P_{\phi}-rQ_{r}+Q=0,
\label{pol2}
\end{equation}
\begin{equation}
m_{\phi}-2r^{2}W_{\phi}+rP\Omega=0,~m_{r}-Q\Omega=0,
\label{pol3}
\end{equation}
\begin{equation}
\frac{\hbar^{2}}{2}\Omega_{\phi}+PW_{r}+\frac{1}{r}QW_{\phi}=0,
\label{pol4}
\end{equation}

\noindent
where we have put $P=k_{1}\cos(\phi)+k_{2}\sin(\phi)$ and 
$Q=-k_{1}\sin(\phi)+k_{2}\cos(\phi)$. 

We see that eq.~\eqref{pol4} contains a term proportional to $\hbar^2$. It 
follows that in this case quantum integrable systems will differ from classical 
ones, at least if we have $\Omega_{\phi}\ne 0$. In the classical limit $\hbar\to 
0$ the quantum systems will reduce to classical ones, or to free motion. This is 
a new phenomenon. In the absence of magnetic fields, classical and quantum 
systems with second order integrals of motion coincide.

Eq.~\eqref{pol1} imply

\begin{equation}
P=-f'(\phi),~Q=f(\phi)+R(r),
\end{equation}

\noindent
with $f(\phi)$ and $R(r)$ to be determined. We shall use primes and dots to 
denote derivatives with respect to $\phi$ and $r$, respectively. We again assume 
$\Omega\ne 0$. Indeed, for $\Omega=0$ we obtain the known case of a scalar 
potential, separable in polar coordinates: $W=W_{0}(r)+\frac{W_{1}(\phi)}{r^2}$.

We solve eq.~\eqref{pol2} for the magnetic field

\begin{equation}
\Omega=-\frac{1}{2r^{3}}(f''+f+R-r\dot{R}).
\label{eqOmega}
\end{equation}

From eq.~\eqref{pol3} we obtain a compatibility condition ($m_{r\phi}=m_{\phi 
r}$), namely

\begin{equation}
2r^{2}W_{r\phi}+4rW_{\phi}+\frac{3f'}{2r^3}(f''+f+R-r\dot{R})+\frac{f'\ddot{R}}
{2r}+\frac{1}{2r^3}(f+R)(f'''+f')=0.
\label{compm}
\end{equation}

Using eq.~\eqref{pol4} to eliminate $W_{\phi}$ from eq.~\eqref{compm}, we 
obtain, for

\begin{equation}
f+R\ne 0,~f'\ne 0
\label{solpart}
\end{equation}

\noindent
an equation for $W(r,\phi)$ that we can solve

\begin{equation}
\begin{array}{ccc}
W_{rr}+\frac{(3(f+R)-r\dot{R})}{r(f+R)}W_{r}
-\frac{\hbar^2\dot{R}(f'''+f')}{4r^3f'(f+R)}
+\frac{3(f+R)}{4r^6}(f''+f+R-r\dot{R})+\frac{(f+R)\ddot{R}}{4r^4}\\
+\frac{(f+R)^2}{4r^6f'}(f'''+f')=0.\\
\end{array}
\end{equation}

We obtain

\begin{equation}
\begin{array}{ccc}
W=\frac{\hbar^2(f'''+f')}{8r^2f'}-\frac{3f(f''+f)}{32r^4}-\frac{f'''}{8f'}r^6 
\dot{u}^2-\frac{r^2}{8}(r^3\ddot{u}+3r^2\dot{u})^2-
\frac{f^2(f'''+f')}{32f'r^4}\\
-\frac{Ff}{2r^2}+(\frac{3f''}{8}+\frac{ff'''}{8f'})r\dot{u}+\frac{3f''}{4}u- 
\frac{f}{4r}(r^3\ddot{u}+3r^2\dot{u})+Fr^3\dot{u}+W_{0},\\
\end{array}
\label{eqW}
\end{equation}

\noindent
where $F=F(\phi)$ and $W_{0}(\phi)$ are two new functions, introduced as 
integration ''constants''. We have also introduced the functions $u(r)$ and 
$S(r)$, satisfying

\begin{equation}
\dot{S}(r)=\frac{1}{r^3}R(r),~\dot{u}(r)=\frac{1}{r^3}S(r)
\end{equation}

Let us first consider the two special cases in eq.~\eqref{solpart}.

For $f(\phi)+R(r)=0$ eq.~\eqref{eqOmega} implies $\Omega=0$ and we are not 
interested in this case.

In the case $f'(\phi)=0$ we have

\begin{equation}
\begin{array}{ccc}
P=0,~Q=Q(r),~W=W(r),~m=\frac{Q^2}{4r^2},\\
k_{1}=-Q(r)\sin(\phi),~k_{2}=Q(r)\cos(\phi),\\
\end{array}
\label{sol1pol}
\end{equation}

\noindent
and the classical and quantum cases coincide. Moreover, a first order integral 
exists.

Let us return to the generic case~\eqref{eqW} with conditions~\eqref{solpart} 
satisfied. We substitute $W$ of eq.~\eqref{eqW} into eq.~\eqref{pol3} 
and~\eqref{pol4} to obtain

\begin{equation}
m=\frac{ff''}{4r^2}+\frac{f^2}{4r^2}+\frac{fR}{2r^2}-\frac{f''S}{2}
+\frac{R^2}{4r^2}+m_{0}(\phi),
\label{eqm}
\end{equation}

\begin{equation}
m_{0}'=2f'F,
\end{equation}

\begin{equation}
\dddot{u}+\frac{6}{r}\ddot{u}
-\frac{r^3\dot{u}^2}{2}(\frac{f'''}{f'})'\frac{1}{f'}+
(\frac{6}{r^2}+\frac{C}{2r^2f'}+\frac{4F'}{f'})\dot{u}+\frac{3f'''}{r^3f'}u+
\frac{4A}{r^7f'}+\frac{4B}{r^5f'}+\frac{4W_{0}'}{r^3f'}=0,\\
\label{equ}
\end{equation}

\noindent
where

\begin{equation}
\begin{array}{ccc}
A=-\frac{9ff'''}{32}-\frac{f^2}{32}(\frac{f'''}{f'})'-\frac{15f'f''}{32}
-\frac{3ff'}{4},\\
B=\frac{\hbar^2}{8}(\frac{f'''}{f'})'-\frac{fF'}{2}-\frac{3f'F}{2},\\
C=6f'''+f(\frac{f'''}{f'})',\\
\end{array}
\end{equation}

The next task is to solve eq.~\eqref{equ}. Notice that this is not a partial 
differential equation. It involves four unknown functions $u(r)$, $f(\phi)$, 
$F(\phi)$ and $W_{0}(\phi)$, each depending on one variable only. Hence, we can 
consider this equation to be an ordinary differential equation for $u(r)$, and 
then establish the compatibility conditions on the other unknowns for which the 
$\phi$ dependence will cancel. The complete analysis is rather lengthy and 
involves the consideration of many special cases. We shall only present the main 
arguments and final results.

Case 1:

\begin{equation}
((\frac{f'''}{f'})'(\frac{1}{f'}))'=0,
\end{equation}

All subcases lead to the following solution (or special cases thereof):

\begin{equation}
\begin{array}{ccc}
f=C_{0}+C_{1}\cos(\phi)+C_{2}\sin(\phi),\\
F=K_{1},~W_{0}=K_{2}f+K_{3}.\\
\end{array}
\end{equation}

Eq.~\eqref{equ} then reduces to

\begin{equation}
\dddot{u}+\frac{6}{r}\ddot{u}+\frac{3}{r^2}\dot{u}-\frac{3}{r^3}u
-\frac{15C_{0}}{8r^7}-\frac{6K_{1}}{r^5}+\frac{4K_{2}}{r^3}=0.
\label{equ2}
\end{equation}

The general solution of eq.~\eqref{equ2} is 

\begin{equation}
u=-\frac{C_{0}}{8r^4}+\frac{2K_{1}}{r^2}+\frac{4K_{2}}{3}+ar+\frac{b}{r}+
\frac{c}{r^3},
\end{equation}

and hence we have

\begin{equation}
\begin{array}{ccc}
S=\frac{C_{0}}{2r^2}-4K_{1}+ar^3-br-\frac{3c}{r},\\
R=-C_{0}+3ar^5-br^3+3cr.\\
\end{array}
\end{equation}

Finally, the magnetic field and effective potential in this case are

\begin{equation}
\Omega=6ar^2-b,
\label{Omegasol2}
\end{equation}

\begin{equation}
W=-2ar(C_{1}\cos(\phi)+C_{2}\sin(\phi))+\frac{ab}{2}r^4-3acr^2-a^2r^6.
\end{equation}

Since $\Omega$ does not depend on $\phi$ the classical and quantum cases are the 
same. The corresponding classical integral of motion is

\begin{equation}
\begin{array}{ccc}
C_{R}=(x\dot{y}-y\dot{x})^2+(-C_{2}-(3ar^5-br^3+3cr)\sin(\phi))\dot{x}\\
+(C_{1}+(3ar^5-br^3+3cr)\cos(\phi))\dot{y}-\frac{3bcr^2}{2}+\frac{9acr^4}{2}\\
-\frac{3abr^6}{2}+2C_{1}ar^3\cos(\phi)-C_{1}br\cos(\phi)+2C_{2}ar^3\sin(\phi)\\
-C_{2}br\sin(\phi)+\frac{9a^2r^8}{4}+\frac{b^2r^4}{4}+\frac{9c^2}{4}.\\
\end{array}
\label{CRsol2}
\end{equation}

Case 2:

\begin{equation}
((\frac{f'''}{f'})'(\frac{1}{f'}))'\ne 0
\end{equation}

A complete analysis~\cite{berube} shown that eq.~\eqref{equ} in this case is 
consistent only if we have

\begin{equation}
u=\frac{a}{8r^4}+\frac{b}{2r^2}+c
\end{equation}

\noindent
and hence

\begin{equation}
S=-\frac{a}{2r^2}-b,~R=a.
\end{equation}

Moreover, the function $f(\phi)$ must satisfy

\begin{equation}
(f+a)^{2}(\frac{f'''}{f'})'+24f'(f+a)+9f'''(f+a)+15f'f''=0.
\label{eqf}
\end{equation}

The functions $F(\phi)$ and $W_{0}(\phi)$ are given explicitly in terms of 
$f(\phi)$ as

\begin{equation}
\begin{array}{ccc}
F=-\frac{bf'''}{4f'}+\frac{\hbar^2}{4(f+a)^3}((f+a)^2\frac{f'''}{f'}
-2(f+a)f''+(f')^2)+\frac{C_{1}}{(f+a)^3},\\
W_{0}=\frac{b^{2}f'''}{8f'}+bF-\frac{3cf''}{4}+C_{2}.\\
\end{array}
\end{equation}

We integrate eq.~\eqref{eqf} twice and put $y=f+a$ to obtain the second order 
equation

\begin{equation}
y''=-\frac{2}{y}(y')^2-3y+\frac{4A}{y}+\frac{B^2-A^2}{y^3}
\end{equation}

\noindent
where $A$ and $B$ are constants.

This equation has a first integral $K$, in terms of which we have

\begin{equation}
y^4(y')^2=-y^6+2Ay^4+(B^2-A^2)y^2+K.
\label{eqy}
\end{equation}

This equation can be written as a quadrature that will express the independent 
variable $\phi$ as a function of $y$ in terms of elliptic integrals. The results 
are not very illuminating, so instead of presenting them, we restrict ourselves 
to some special cases. Let us first rewrite eq.~\eqref{eqy} as 

\begin{equation}
y^4(y')^2=-(y^2-y_{1}^2)(y^2-y_{2}^2)(y^2-y_{3}^2)\equiv T(y),
\label{eqT}
\end{equation}

\noindent
where the roots $y_{1}$, $y_{2}$ and $y_{3}$ are related to the constants $A$, 
$B$ and $K$ by the formulas

\begin{equation}
K=y_{1}^2y_{2}^2y_{3}^2,~B^2-A^2=-(y_{1}^2y_{2}^2+y_{2}^2y_{3}^2+y_{3}^2y_{1}^2)
,~2A=y_{1}^2+y_{2}^2+y_{3}^2.
\end{equation}

\begin{figure}[htp]
  \begin{center}
    \subfigure[Three pairs of simple roots]{\label{fig1a} 
    \includegraphics[height=7cm,
      width=6.5cm]{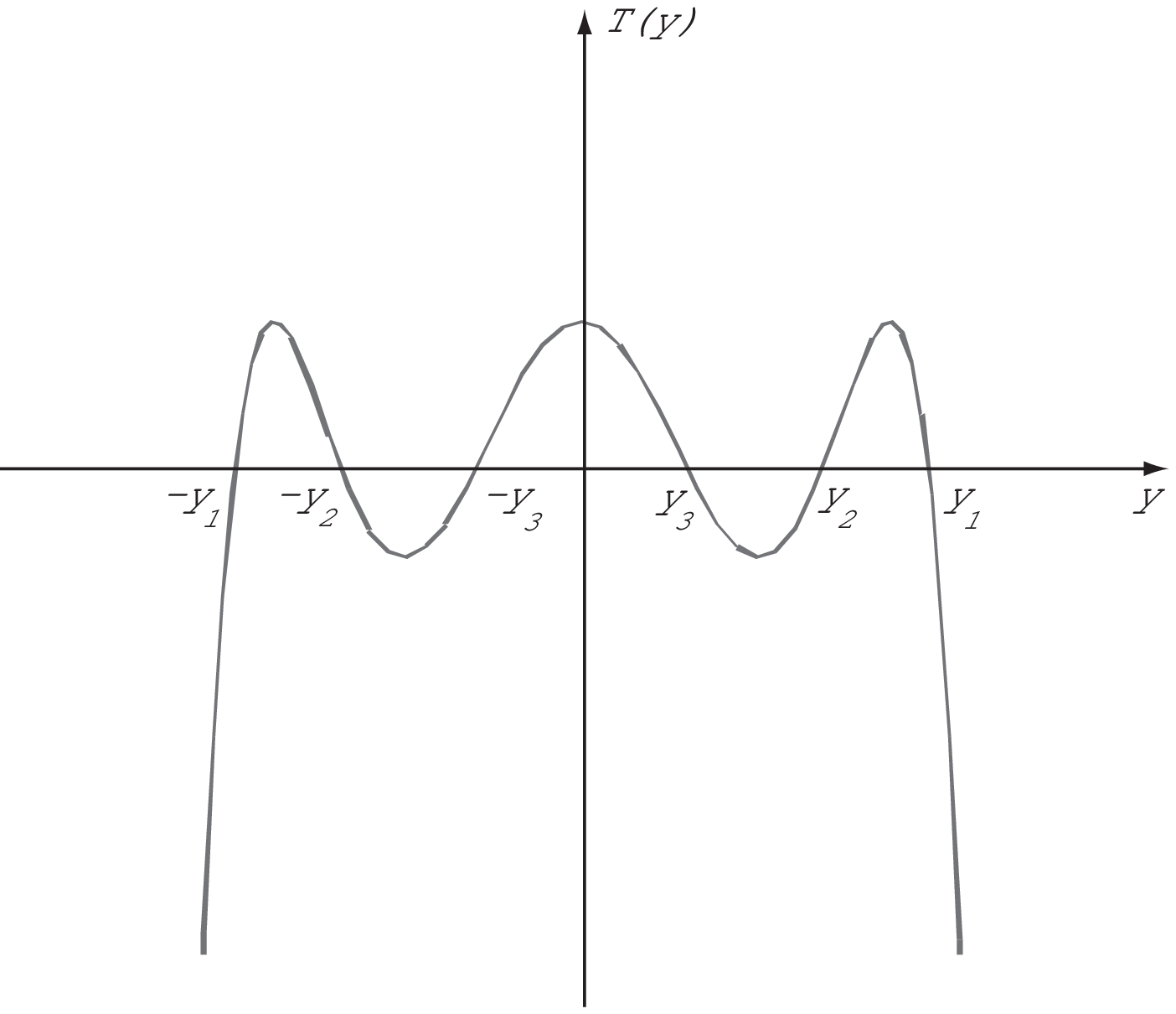}}
    \subfigure[One quadruple root, a pair of single ones]{\label{fig1b} 
\includegraphics[height=7cm,
      width=6.5cm]{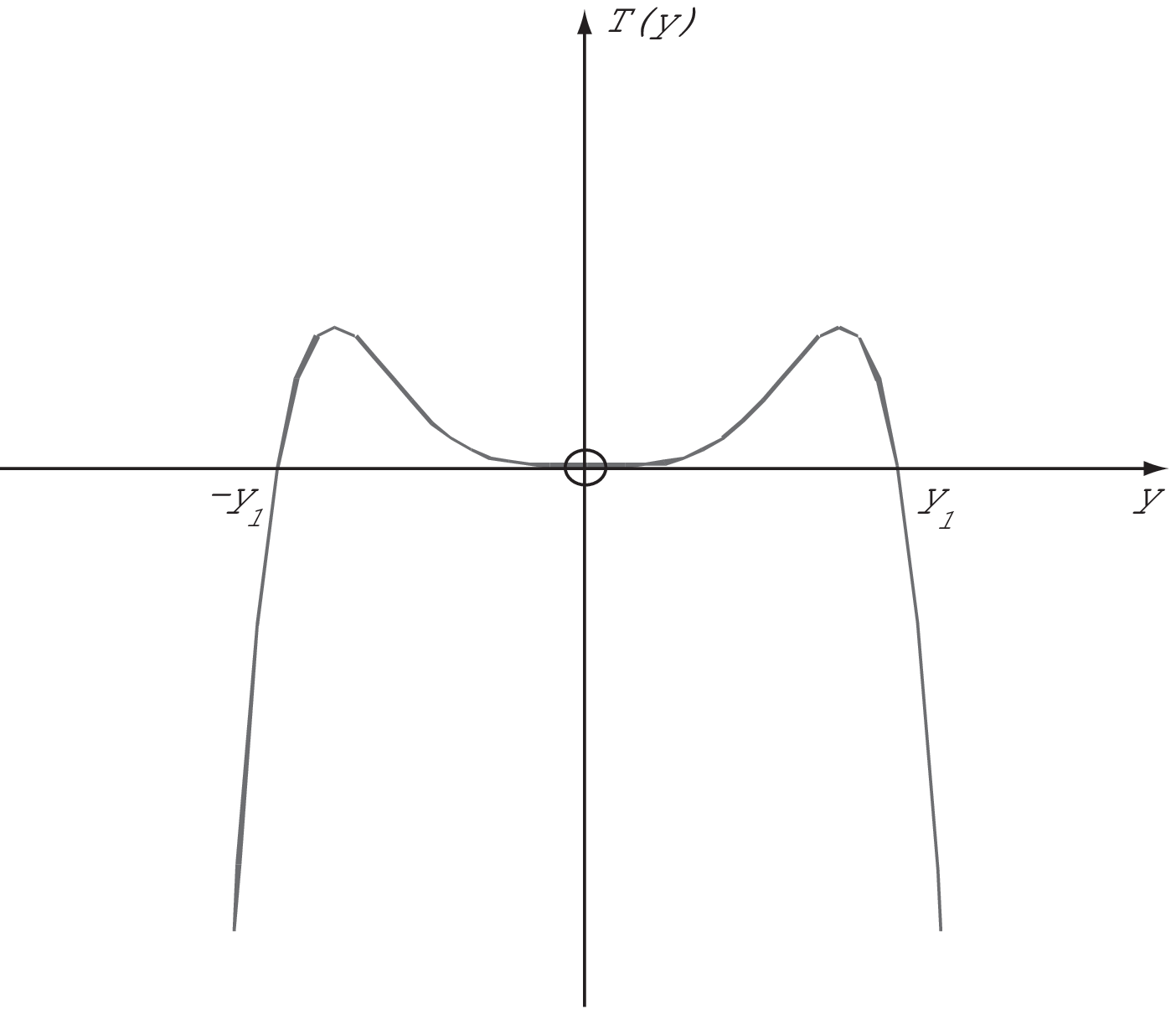}}
     \subfigure[One double root, two pairs of single ones]{\label{fig1c} 
    \includegraphics[height=7cm,
      width=6.5cm]{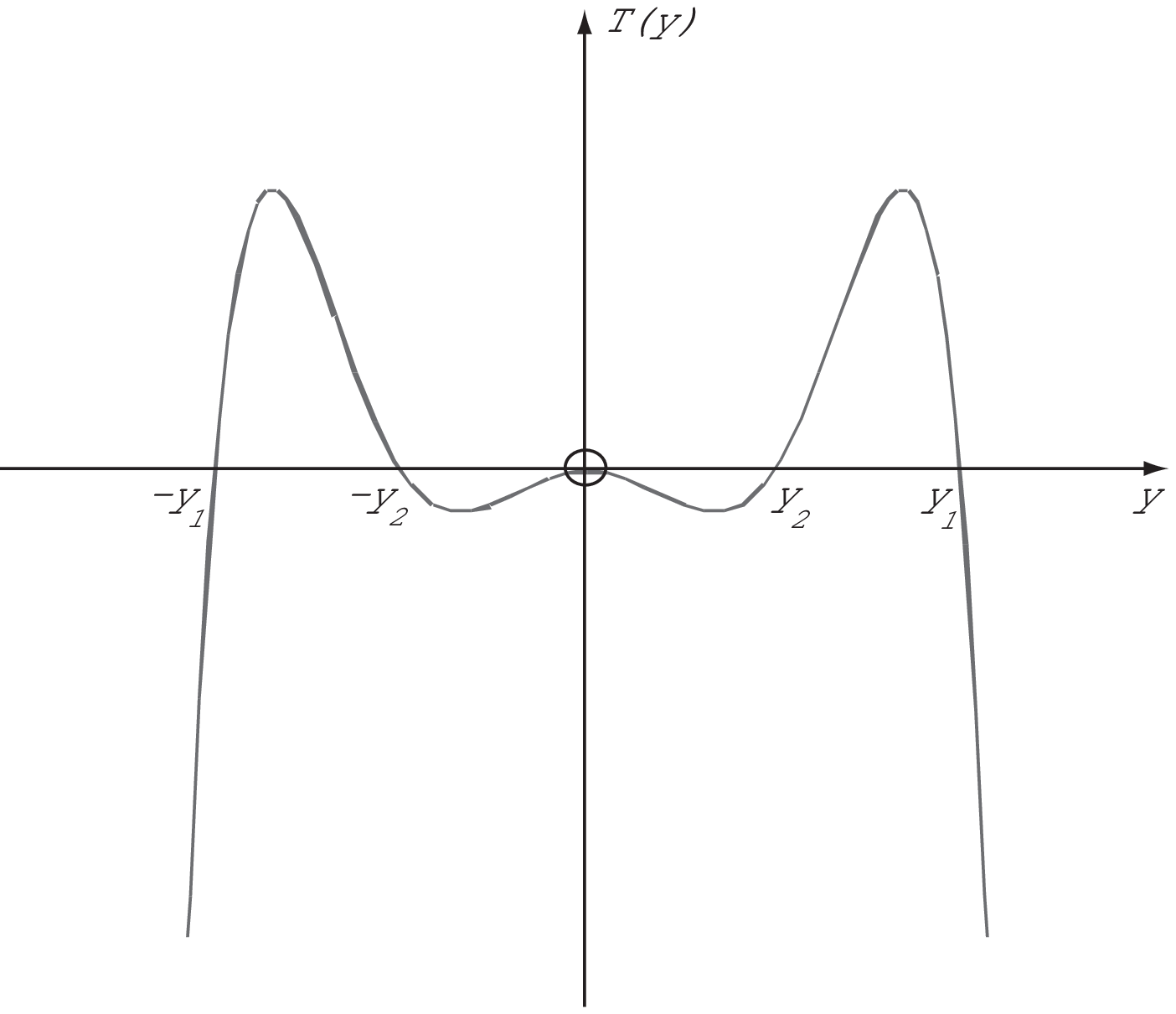}}
      \subfigure[One pair of double roots and one of simple ones]{\label{fig1d} 
    \includegraphics[height=7cm,
      width=6.5cm]{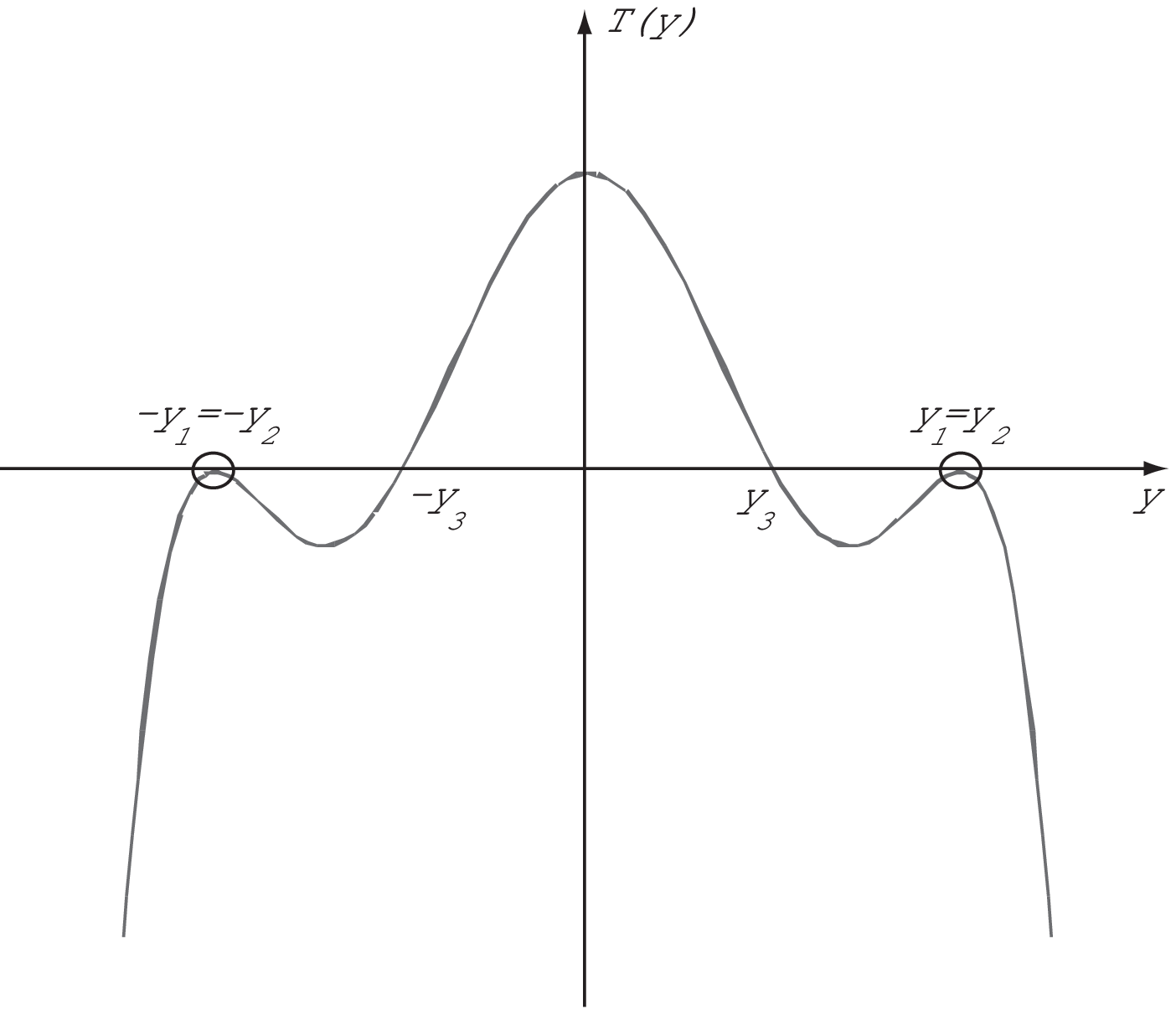}}
  \caption{Roots of the polynomial T(y) in eq.~\eqref{eqT}}
      \end{center}
\end{figure}

If all the roots $y_{i}$ are real, the behavior of the polynomial $T(y)$ as a 
function of $y$ is shown on Fig.~\ref{fig1a}.

If all roots are distinct ($0<y_{3}<y_{2}<y_{1}<\infty$), real periodic 
solutions are obtained for $-y_{3}\leq y\leq y_{3},~y_{2}\leq y\leq y_{1}$ and 
$-y_{1}\leq y\leq y_{2}$. However, these are expressed in 
terms of elliptic functions and the period is not a multiple of $\pi$. Constant 
solutions of eq.~\eqref{eqT} are obviously $y=\pm y_{k}$, $k=1,2~or~3$.

Elementary $\phi$ dependent real finite periodic solutions are obtained whenever 
the polynomial $T(y)$ has multiple roots. The corresponding solutions are

(1) $y_{3}=y_{2}=0$, $y_{1}>0$ (See Fig.~\ref{fig1b})

\begin{equation}
y=y_{1}\sin(\phi-\phi_{0})
\end{equation}

(2) $0=y_{3}<y_{2}<y_{1}$ (See Fig.~\ref{fig1c})

\begin{equation}
y=\pm\frac{1}{\sqrt{2}}\sqrt{y_{1}^2+y_{2}^2+(y_{1}^2-y_{2}^2)
\sin 2(\phi-\phi_{0})}
\end{equation}

or in terms of $A$ and $B$:

\begin{equation}
y=\pm\sqrt{A+B\sin 2(\phi-\phi_{0})}
\end{equation}

(3) $0<y_{3}<y_{2}=y_{1}$ (See Fig.~\ref{fig1d})

In this case we give the solution $y$ implicitly as

\begin{equation}
\begin{array}{ccc}
-2\sqrt{y_{1}^2-y_{3}^2}\arcsin(\frac{y}{y^3})+y_{1}(\arcsin(
\frac{y_{3}^2+yy_{1}}{y_{3}(y+y_{1})}\\
-\arcsin(\frac{y_{3}^2-yy_{1}}{y_{3}(y-y_{1})})=\pm 
2\sqrt{y_{1}^2-y_{3}^2}(\phi-\phi_{0})
\end{array}
\end{equation}

The solution is real, finite and periodic for $-y_{3}\leq y\leq y_{3}$.

For any solution $y(\phi)$ of eq.~\eqref{eqy} we obtain a magnetic field and 
effective potential in the form

\begin{equation}
\Omega=-\frac{f''+f+a}{2r^3}
\label{Omegasol3}
\end{equation}

\begin{equation}
W=\frac{\hbar^2}{8r^2}(1+\frac{2f''}{f+a}-\frac{(f')^2}{(f+a)^2})
-\frac{(f+a)^2}{32r^4}(\frac{f'''}{f'}+4)-\frac{3f''(f+a)}{32r^4}
-\frac{C_{1}}{2r^2(f+a)^2}+C_{2},
\end{equation}

The functions $P$, $Q$ and $m$ figuring in the polar integral are

\begin{equation}
\begin{array}{ccc}
P=-f'(\phi),~Q=f(\phi)+a,\\
m=\frac{ff''+(f+a)^2+af''}{4r^2}-\frac{bf''}{2}-\frac{C_{1}}{(f+a)^2}+
\frac{\hbar^2}{4(f+a)^2}(2(f+a)f''-(f')^2).
\end{array}
\label{PQsol3}
\end{equation}

Let us sum up the results of this section. Three different cases of polar 
integrability exist. They are given by eq.~\eqref{sol1pol},~\eqref{Omegasol2} 
to~\eqref{CRsol2} and~\eqref{Omegasol3} to~\eqref{PQsol3}, respectively. The 
last case provides an example where the quantum system and the classical one 
differ. Indeed, the Planck constant figures explicitly in the effective 
potential $W$ and in the integral of motion.

\section[]{Polar superintegrability}

Let us assume that we have a Hamiltonian~\eqref{eqH} that is ``polar 
integrable'', i.e. allows an integral of motion of the form $X_{R}$ as in 
eq.~\eqref{Xpol}. The magnetic field $\Omega$ and effective potential $W$ must 
hence have one of the three forms established in Section 5. For the system to be 
superintegrable, it must allow at least one further integral, by assumption of 
the form~\eqref{opeX}. We can simplify this second integral by linear 
combinations with $X_{R}$ and with $H$ and also by rotations, since they will 
not destroy the form of $X_{R}$ (nor $H$). Thus, in eq.~\eqref{opeX} we take 
$\alpha=0$, $\zeta=-\delta$. Furthermore, we can assume $\beta^2+\gamma^2\ne 0$, 
since otherwise we would be in the case of cartesian superintegrability, already 
treated in Section 4. By a rotation and normalization, we can set $\beta=1$, 
$\gamma=0$. It follows that the second integral $X_{2}$ is of the parabolic 
type, conjugate to $X_{P}$ of eq.~\eqref{Xpar}.

The determining equations for $X_{2}$, obtained from~\eqref{eq2345} 
and~\eqref{eq6} are

\begin{equation}
k_{1,x}-(x+\xi)\Omega=0,~k_{2,y}+(x+\xi)\Omega=0,
\label{super1}
\end{equation}

\begin{equation}
-2\beta\Omega y+k_{1,y}+k_{2,x}=0,
\label{super2}
\end{equation}

\begin{equation}
\begin{array}{ccc}
-2yW_{x}+(x+\xi)W_{y}+k_{2}\Omega-m_{x}=0,\\
(x+\xi)W_{x}-k_{1}\Omega-m_{y}=0,\\
\end{array}
\label{super3}
\end{equation}

\begin{equation}
k_{1}W_{x}+k_{2}W_{y}+\frac{\hbar^2}{4}\beta\Omega_{x}=0.
\label{super4}
\end{equation}

For each of the three polar integrable systems we obtain the same result, 
namely: equations~\eqref{super1} to~\eqref{super4} are compatible only if 
$\Omega=\Omega_{0}$ and $W=W_{0}$ are constant. Then we have three first order 
integrals and the corresponding second order integrals are polynomial in the 
first order ones.

\section[]{Conclusions}

We have constructed all integrable quantum systems with a vector and scalar 
potential (as in eq.~\eqref{eqH}) that possess either a first order integral, or 
a second order one of the cartesian, or polar type.

It is interesting to compare such systems with a nonzero magnetic field $\Omega$ 
with systems allowing a scalar potential only.

1. The first difference is that for $\Omega\ne 0$ quantum and classical 
integrable systems with second order integrals do not necessarily coincide. The 
Planck constant $\hbar$ can figure in a nontrivial way in the potentials and 
integrals of motion.

2. The existence of a first order integral of motion implies a geometrical 
symmetry, both for $\Omega\ne 0$ and $\Omega=0$. Indeed, a first order integral 
exists if and only if we have either $\Omega=\Omega(r)$, $W=W(r)$, or 
$\Omega=\Omega(y)$, $W=W(y)$ (up to Euclidean transformations). The functions 
$\Omega$ and $W$ are arbitrary in both cases.

3. The existence of a second order integral for $\Omega=0$ implies that the 
Schrödinger equation will allow separation of variables in cartesian, polar, 
parabolic, or elliptic coordinates. In each case the potential $V(x,y)$ depends 
on two arbitrary functions of one variable. For $\Omega\ne 0$ the coordinates no 
longer separate. The requirement that an irreducible second order integral 
should exist for $\Omega\ne 0$ is much more restrictive than for $\Omega=0$. The 
quanttities $\Omega(x,y)$ and $W(x,y)$ again depend on two functions of one 
variable, however these functions obey certain ordinary differential equations. 
They are hence determined completely, up to some arbitrary constants. For 
instance, in the cartesian case, they are elliptic functions, or degenerate 
cases of elliptic functions.

4. For $\Omega=0$ four families of superintegrable systems in $E(2)$ 
exist~\cite{fris1}, each depending on three parameters. For $\Omega\ne 0$ we 
have shown that superintegrability with first order integrals of the cartesian, 
or polar type, exists only for $\Omega$ and $W$ constant.

Several related problems are presently under consideration. To complete the 
study of quadratic integrability in $E(2)$ for $\Omega\ne 0$ we must still 
consider parabolic and elliptic integrability. For $\Omega=0$ their is a close 
relation between superintegrability and exact solvability~\cite{turb}. For 
$\Omega\ne 0$ the requirement of superintegrability seems to be too restrictive. 
An important question is whether some of the integrable systems found in this 
article are actually exactly solvable.

\section*{Acknowledgements}

The research of P.W. was partly supported by research grants from NSERC of 
Canada and FQRNT du Québec.

\end{document}